# Interlayer Exchange Coupling-Induced Critical-Metal-to-Insulator Phase Transition in Quantum Anomalous Hall Insulators

Ruoxi Zhang[1,3], Yi-Fan Zhao[1,3], Ling-Jie Zhou[1], Deyi Zhuo[1], Zi-Jie Yan[1], Chao-Xing Liu[1], Moses H. W. Chan[1], Chui-Zhen Chen[2], and Cui-Zu Chang[1]

[1] Department of Physics, The Pennsylvania State University, University Park, PA 16802, USA

[2] Institute for Advanced Study and School of Physical Science and Technology, Soochow University, Suzhou 215006, China

[3] These authors contributed equally: Ruoxi Zhang and Yi-Fan Zhao

Corresponding authors: cxc955@psu.edu (C.-Z. Cha.); czchen@suda.edu.cn (C.-Z. Che.).

**Abstract: Interlayer exchange coupling (IEC) between two magnetic layers sandwiched by a nonmagnetic spacer layer plays a critical role in shaping the magnetic properties of such heterostructures. The quantum anomalous Hall (QAH) effect has been realized in a structure composed of two magnetically doped topological insulator (TI) layers separated by an undoped TI layer. In this work, we employ molecular beam epitaxy to synthesize a series of magnetic TI sandwiches with varying thicknesses of the middle TI spacer layer. The well-quantized QAH effect is observed in all these samples, and the IEC modulates its critical behavior between the top and bottom magnetic TI layers. Near the plateau phase transition (PPT), thinner QAH samples exhibit a two-dimensional critical metal behavior with nearly temperature-independent longitudinal resistance. In contrast, thicker QAH samples behave as a three-dimensional insulator with reduced longitudinal resistance at higher temperatures. We employ a magnetic TI Hamiltonian with random magnetic domains to understand the IEC-induced critical-metal-to-insulator transition observed near QAH PPT.**

**Main text:** When a nonmagnetic layer is sandwiched between two magnetic layers, the magnetization of these two magnetic layers interacts through interlayer exchange coupling (IEC). The strength of the IEC can be tuned by varying the thickness of the nonmagnetic spacer layer. With a metallic spacer layer, the IEC usually exhibits a power-law decay with oscillations as a function of the spacer layer thickness due to the Ruderman-Kittel-Kasuya-Yosida (RKKY) interaction via itinerant carriers [1,2]. In contrast, with an insulating spacer layer, the IEC usually decays exponentially with the spacer layer thickness [3,4] due to Bloembergen-Rowland interaction [5]. As a new type of insulator, topological insulator (TI), in which the bulk interior is insulating but its surfaces/edges are metallic [6,7], has recently been employed as a spacer layer in magnetic TI/TI/magnetic TI sandwiches to realize the quantum anomalous Hall (QAH) effect [8-11].

The QAH effect is a zero magnetic field manifestation of the quantum Hall (QH) effect [8]. Unlike the QH effect, which relies on the formation of Landau levels in a two-dimensional (2D) electron gas under high magnetic fields [12-15], the QAH effect is realized as a result of the intrinsic property of electronic band structures and broken time-reversal symmetry [8,16-23]. The QAH effect is first experimentally realized in magnetically doped TI, specifically Cr-doped $(Bi,Sb)_2Te_3$ films [8,24-26]. In the QH effect, the plateau phase transition (PPT) corresponds to the quantum state that exhibits localization to delocalization and back to localization [15,27-29]. In contrast, the PPT in the QAH effect corresponds to the quantum state wherein a single magnetic domain gives way to multiple magnetic domains and then re-converges back to a single magnetic domain [8]. However, the influence of magnetic domains on the PPT remains inadequately understood, primarily due to the lack of effective knobs to regulate these domains. So far, the structure for achieving the QAH effect with a higher critical temperature is the magnetic TI/TI/magnetic TI sandwich because the quantized Hall conductance in QAH insulators is a combined contribution from the top and bottom surfaces [8-11]. Prior studies [8-11,30-33] have confirmed the ferromagnetic IEC in magnetic TI sandwiches and a monotonic decrease in IEC strength with increasing thickness of the TI spacer layer. These magnetic TI sandwiches with the QAH state can serve as a testbed for exploring the role of the TI spacer layer and the impact of IEC on the QAH effect.

In this work, we employ molecular beam epitaxy (MBE) to synthesize a series of magnetic TI sandwiches, i.e., 3 quintuple layers (QLs) $(Bi,Sb)_{1.76}Cr_{0.24}Te_3$/$m$QL $(Bi,Sb)_2Te_3$/3QL $(Bi,Sb)_{1.76}Cr_{0.24}Te_3$ with $0 \leq m \leq 100$ (Fig.1a). All these samples exhibit the well-quantized QAH effect. We find that the maximum values of their longitudinal resistance $\rho_{xx,\max}$ near PPT show a systematic behavior with increasing $m$. Thinner QAH samples exhibit 2D critical metal behaviors, with their $\rho_{xx,\max}$ being nearly independent of temperature. In contrast, thicker QAH samples behave as 3D insulators, with their $\rho_{xx,\max}$ decreasing with temperature. The IEC-induced critical-metal-to-insulator transition (CMIT) near QAH PPT can be understood using a magnetic TI Hamiltonian with random magnetic domains. Our measurements demonstrate the critical role of IEC in controlling the magnetic domains and quantum phase transition near QAH PPT.

We first perform electrical transport measurements on all magnetic TI sandwiches at charge neutral points $V_g=V_g^0$ and $T$=25mK. The value of $V_g^0$ is determined when the zero magnetic field Hall resistance $\rho_{yx}(0)$ is maximized. For $m$=0, i.e., a 6QL Cr-doped $(Bi,Sb)_2Te_3$ film, $\rho_{yx}(0) \sim 0.873h/e^2$, concomitant with the zero magnetic field longitudinal resistance $\rho_{xx}(0) \sim 0.509h/e^2$. Note that the value of $\rho_{yx}(0)/\rho_{xx}(0)$ is greater than 1, indicating the presence of the QAH state in the $m = 0$ sandwich [8]. In magnetically doped TI films, besides introducing the ferromagnetic order, Cr doping also reduces the spin-orbit coupling of the magnetic TI layer and drives its bulk energy gap towards the normal insulator regime [8,11,34-36]. Consequently, the quantization quality of the individual magnetically doped TI films is highly dependent on the Cr doping concentration $x$. For the $m$=0 sample, i.e., an individual 6 QL $(Bi,Sb)_{1.76}Cr_{0.24}Te_3$ film, the Cr doping concentration in the top and bottom Cr-doped TI layer is $x$~0.24, which is much higher than the Cr doping concentration of the individual magnetically doped TI films and within the nontrivial-to-trivial phase transition region [8,11,34]. Therefore, the $m$=0 sample exhibits an imperfect QAH state. For $1 \leq m \leq 100$, the observations of quantized $\rho_{yx}(0)$ and vanishing $\rho_{xx}(0)$ confirm the perfect QAH effect (Figs.1b to 1d, S1 to S3, and Table S1) [37]. The well-quantized QAH state is further validated by the quantized $\rho_{yx}(0)$ and zero $\rho_{xx}(0)$ plateaus in the gate-dependent measurements (Fig. S1) [37]. The observation of the well-quantized QAH state in thick magnetic TI samples implies that the thickness $m$ of the insulating undoped TI spacer layer has virtually no impact on the QAH behaviors, consistent with our recent studies [33,43].

Despite the consistent QAH behaviors within the well-defined magnetization regime, a systematic dependence on $m$ is observed in $\rho_{xx,\max}$ near PPT, specifically near the coercive field $\mu_0 H_c$. To elucidate this behavior, we define an effective magnetic field $\mu_0 H_E = \mu_0(H-H_c)$ and plot $\rho_{xx}$–$\mu_0 H_E$ curves of the sandwiches with $4 \leq m \leq 100$ near PPT (Figs. 1c and 1d). For $m \leq 16$, $\rho_{xx,\max}$ is found to be ~$h/e^2$. For $m$>16, $\rho_{xx,\max}$ increases with $m$ gradually and then more abruptly between $m$=18 and $m$=20. The values of $\rho_{xx,\max}$ are ~$1.155h/e^2$, ~$3.778h/e^2$, and ~$18.734h/e^2$ for the $m$ = 16, 18, and 20 samples, respectively (Figs. 1c and 1e). With an increase in $m$, $\rho_{xx,\max}$ decrease rapidly. The values of $\rho_{xx,\max}$ are ~$3.855h/e^2$, ~$1.757h/e^2$, and ~$1.238h/e^2$ for the $m$=30, $m$=50, and $m$=100 samples, respectively (Figs. 1d and 1e). $\rho_{xx,\max}$ shows a sharp peak at $m$=20 (Fig. 1e). We note that the large $\rho_{xx,\max}$ behavior has been observed in a second batch of the QAH sandwiches with $m$=17 and $m$=20 (Fig. S6).

Next, we convert $\rho_{yx}$ and $\rho_{xx}$ of the sandwiches with 4≤$m$≤100 into Hall conductance $\sigma_{xy}$ and longitudinal conductance $\sigma_{xx}$. For $m$≤16, both $\sigma_{xy}$ and $\sigma_{xx}$ exhibit a one-step transition during PPT (Figs. 2a to 2c). For $m$=18, a two-step transition emerges in both $\sigma_{xy}$ and $\sigma_{xx}$ (Fig. 2d). For $m$=20, a sharp $\sigma_{xx}$ dip towards zero is observed at ~$\mu_0 H_c$, corresponding to a narrow $\sigma_{xy}$=0 plateau (Fig. 2e). With a further increase in $m$, the two-step transition feature weakens (Figs. 2f to 2h). For $m$=100, the two-step transition feature nearly disappears in $\sigma_{xy}$, but a double peak feature remains observable in $\sigma_{xx}$ (Fig. 2h). Note that both $\sigma_{xx}$ and $\rho_{xx}$ show kink features after the magnetic field crosses the zero magnetic field (Figs. 2 and S1). This kink feature is likely related to temperature rise in the sample induced by magnetic field polarity reversal during the magnetic field sweep [34,44]. The evolution of the QAH transport behaviors near PPT becomes much clearer by plotting their flow diagrams ($\sigma_{xy}$, $\sigma_{xx}$) (Figs. 2i to 2p). As $m$ increases, the flow diagram ($\sigma_{xy}$, $\sigma_{xx}$) transforms from a single semicircle centered at ($\sigma_{xy}$, $\sigma_{xx}$) = (0, 0) for $m$=4 to two semicircles centered at ($\sigma_{xy}$, $\sigma_{xx}$)=($\pm e^2/2h$, 0) for $m$=20 and then gradually reverts to a single semicircle centered at ($\sigma_{xy}$, $\sigma_{xx}$)=(0, 0) for $m$=100. The deviation from the standard one- or two-semicircle behavior can arise from different physical mechanisms depending on the range of $m$. For $m$=4 and $m$=8, the slight deviation from the one semicircle might result from imperfections in the local area of the sample (Figs. 2i and 2j). For $m \geq 30$, the deviation from the two semicircle behavior is likely to be attributed to the emergence of conduction channels from side surface states and the bulk (Figs. 2n to 2p) [33,43]. However, for $16 \leq m \leq 20$, the transition from one- to two-semicircle behavior in the flow diagram may indicate a quantum phase transition as $m$ increases (Figs. 2k to 2m).

The appearance of a two-semicircle feature in ($\sigma_{xy}$, $\sigma_{xx}$) results from the presence of the $\sigma_{xy}$=0 plateau, which can originate from two different physical mechanisms [8]: (*i*) the presence of the hybridization gap in thinner uniformly doped magnetic TI films [45-48]; and (*ii*) independent magnetic reversal on top and bottom surfaces in thicker magnetic TI sandwiches [31,32]. In uniformly doped magnetic TI films, the two-semicircle feature in ($\sigma_{xy}$, $\sigma_{xx}$) is considered a hallmark of 2D QAH insulators [45,48]. However, this criterion does not apply to magnetic TI sandwiches [8,31-33,49]. In our experiments, the observation of one semicircle in $m$=4 and $m$=8 samples indicates that all QAH sandwiches are in the 3D QAH state regime [43]. Therefore, the nonmonotonic evolution of the QAH flow diagram also suggests that the two-semicircle feature in ($\sigma_{xy}$, $\sigma_{xx}$) for $m$=20 may arise from a different physical origin. As noted above, this observation in thicker magnetic TI sandwiches is typically associated with decoupled magnetization reversal between the top and bottom magnetic TI layers. This decoupled magnetization reversal behavior is a direct consequence of the weakening of IEC for large $m$, making the two-step transition in the flow diagrams a signature of reduced IEC strength in our QAH sandwiches.

The observation of the two-step transition in the flow diagrams for $16 \leq m \leq 20$ at $T = 25$ mK reflects the emergence of an intermediate regime with an enhanced longitudinal resistance ($\rho_{xx,\max} \gg h/e^2$). However, the above observation at $T = 25$ mK alone is insufficient to determine the nature of the intermediate regime. To examine the potential quantum phase transition and uncover the underlying physics behind the large $\rho_{xx,\max}$ and the two-step transition in QAH sandwiches with $16 \leq m \leq 20$, we investigate the temperature dependence of $\rho_{xx}$ near PPT with varying $m$. In ferromagnetic materials, $\mu_0 H_c$ typically increases as $T$ decreases, so $\rho_{xx,\max}$ appears at different $\mu_0 H$ values under different $T$. For $m$=4 and $m$=8, the value of $\rho_{xx,\max}$ remains nearly constant ~$h/e^2$ for $T$≤350mK. Upon raising $T$ to ~500mK, the value of $\rho_{xx,\max}$ increases slightly to ~1.2 $h/e^2$ and retains this value up to $T$=1K (Figs. 3a, 3e, and S3) [37]. This behavior is independent of changes in ferromagnetic order with temperature, as indicated by $\mu_0 H_c$ as the position of $\rho_{xx,\max}$.

This nearly temperature-independent $\rho_{xx,\max}$ observed for 25 mK ≤ $T$ ≤ 350 mK suggests a critical metal (CM) behavior near PPT. The CM phase can be understood as a line of critical points with temperature-dependent behavior between conventional metal and insulator states [50-52]. A single critical point has been observed in the QH to Hall insulator transition [53-55] and the QAH to Anderson insulator transition [56,57]. Nevertheless, prior studies have never observed the CM phase despite its prediction in numerous theoretical works [50,52,58-63]. Our work provides concrete experimental evidence of the CM phase. We note that in a prior study [57], a temperature-independent $\rho_{xx,\max}$ behavior is also observed for $T$ ≥ 300 mK. However, in the magnetically doped QAH samples, $\rho_{xx,\max}$ often remains relatively constant at higher temperatures but begins to increase or decrease again as $T$ further decreases. For example, the $m$ = 50 and $m$ = 100 samples show $\rho_{xx,\max}$ ~ $h/e^2$ for 500 mK ≤ $T$ ≤ 1 K, but $\rho_{xx,\max}$ increases suddenly for $T$ ≤ 500 mK (Fig. 3d and S4c). Therefore, the temperature-independent $\rho_{xx,\max}$ behavior observed for $T$ ≥ 300 mK [57] is insufficient to confirm the presence of the CM phase. In contrast, the temperature-independent $\rho_{xx,\max}$ behavior observed in this work for 25 mK ≤ $T$ ≤ 350 mK demonstrates that the CM phase is a valid ground state.

For $m$>16, the value of $\rho_{xx,\max}$ consistently decreases as $T$ increases from 25mK to 1K, showing an insulating behavior (Figs. 3c, 3d, 3e, 3f, and S4) [37]. For $m$=16, the value of $\rho_{xx,\max}$ stabilizes at ~1.15$h/e^2$ for 25mK≤$T$≤250mK and shows a slight increase to ~1.19$h/e^2$ at $T$=350mK. With further increasing $T$, $\rho_{xx,\max}$ gradually decreases to ~1.04$h/e^2$ at $T$=1K (Fig. 3b and 3e). This $\rho_{xx,\max}$-$T$ curve indicates that the $m$=16 sandwich is in an intermediate state, exhibiting a CM behavior for 25mK≤$T$≤350mK and transitioning to an insulating behavior for $T$≥350mK. In addition, we find that $\rho_{xx,\max}$-$T$ relationship in the insulating phase follows the variable range hopping model of Anderson localized states, described by $1/\rho_{xx,\max} \propto \exp[-(T_0/T)^{\beta}]$ with β=0.5 (Fig. S9) [64]. Therefore, we observe a CMIT near QAH PPT induced by an increase in $m$. We note that the discrepancy between the mixing chamber temperature and the electron temperature for $T$ < 100 mK [37] does not alter the overall trends of the temperature-dependent $\rho_{xx,\max}$ curves in Figs. 3e and 3f.

Based on the above discussions, the weakening of IEC with increasing $m$ serves as a common origin for both the emergence of the two-step transition in the flow diagrams and the occurrence of the CMIT. Although the two-step transition in the flow diagrams corresponds to the magnetization dynamics at base temperature (i.e., $T$ = 25 mK) and reflects the decoupling of the magnetizations in the top and bottom magnetic TI layers, the $T$ scaling of $\rho_{xx,\max}$ reveals the quantum critical nature of the intermediate state. These complementary observations demonstrate that varying $m$ enables control over the IEC strength, which in turn drives both the CMIT and the two-step transition in the flow diagram.

In a magnetically doped TI, a massive Dirac-Hamiltonian can describe the top and bottom surface states [8]. In each uniform magnetic domain, its Chern number $C$ depends on the magnetization alignment between the top and bottom surface magnetic layers. Each surface contributes a Hall conductance of $(m_z/|m_z|)e^2/2h$, where $m_z$ is the surface magnetization [16]. The total $\sigma_{xy}$ is composed of the joint contribution of the magnetic layers on the top and bottom of the surface. With a parallel magnetization alignment, $\sigma_{xy}$~$\pm(e^2/2h + e^2/2h)$ and the magnetic domains behave as the $C$=±1 QAH effect with one chiral edge channel [8]. In contrast, with an antiparallel magnetization alignment, $\sigma_{xy}$~0, the magnetic domains behave as an axion insulator without the formation of the chiral edge channel (Fig. 4a) [8,31,32]. Our local Chern marker calculations [65,66] confirm that the percentage of the antiparallel magnetization domains (i.e., $C$=0 magnetic domains and $\sigma_{xy}$~0) increases as the percentage of antiparallel magnetization states increases (Figs. 4e, 4f, S7, and S8) [37].

In our magnetic TI sandwiches, both the top and bottom surface layers are Cr-doped, and their magnetizations can be easily parallel aligned, guaranteeing the appearance of the QAH effect as long as the middle TI spacer layer remains insulating [43]. However, near PPT, multiple small magnetic domains with random magnetization directions are formed in the top and bottom magnetic TI layers. Therefore, the local magnetization alignment between the top and bottom magnetic TI layers can be influenced by varying $m$. For thinner QAH sandwiches, the stronger IEC promotes the parallel magnetization alignment within magnetic domains, and thus, the chiral edge state propagates along the edges of each magnetic domain [31-33]. This configuration favors the tunneling among chiral edge modes surrounding magnetic domains (Fig. 4b). However, for thicker QAH sandwiches, weaker IEC is not able to prevent antiparallel magnetization alignment between multiple magnetic domains in the top and bottom magnetic TI layers, resulting in insulating domains that behave as axion insulators without chiral edge states. This configuration impedes the tunneling among chiral edge modes surrounding magnetic domains. It thus increases the value of $\rho_{xx,\max}$ near PPT for $m$>16 (Fig. 4c). Therefore, weaker IEC increases the number of $C$=0 magnetic domains, which in turn impedes the percolation of chiral edge modes, leading to the occurrence of CMIT in our QAH sandwiches. A slight difference of $\mu_0 H_c$ between the top and bottom magnetic TI layers may also exist due to different chemical environments, further enlarging the regions of the $C$=0 magnetic domains [67]. The decrease in $\rho_{xx,\max}$ values for $m \geq 20$ is likely a result of the side surface carriers in thicker QAH samples [33,43].

To quantitatively understand the IEC-induced CMIT near QAH PPT, we perform the standard finite-size scaling analysis based on a magnetic TI Hamiltonian with random magnetic domains [61]. We begin with a 4×4 effective Hamiltonian of a magnetic TI and introduce magnetic domains with random (±) magnetization alignment between the top and bottom magnetic TI layers [61]. The effective Hamiltonian can be written as:

$$H_{\mathrm{TI}}(\boldsymbol{k}) = \begin{pmatrix} H_{\mathrm{t}}(\boldsymbol{k}) & m_{\boldsymbol{k}} \\ m_{\boldsymbol{k}} & H_{\mathrm{b}}(\boldsymbol{k}) \end{pmatrix} + h_{\mathrm{z}}(\boldsymbol{r}) \qquad (1)$$

where $H_{\mathrm{t/b}} = \pm \hbar v_{\mathrm{F}} (s_x k_y - s_y k_x)$ denotes the Dirac Hamiltonian for the top/bottom surface states with $v_{\mathrm{F}}$ being the Fermi velocity. The Pauli matrices $s_{x,y,z}$ act on spin space, *t/b* denotes the top/bottom surface of the TI, and $k_{x,y}$ are wave vectors. The term $m_{\boldsymbol{k}} = m_0 + m_1 k^2$ describes the hybridization gap between the top and bottom surfaces. The spatially dependent exchange field is given by $h_{\mathrm{z}}(\boldsymbol{r}) = m_{\mathrm{z}}(\mathrm{r}) s_z$ when the magnetic domains have parallel magnetization alignment along $\pm z$ between the top and bottom magnetic TI layers. However, $h_{\mathrm{z}}(\boldsymbol{r}) = m_{\mathrm{z}}(\mathrm{r}) s_z \tau_z$ with Pauli matrix $\tau_z$ acting on the *t/b* space, indicates antiparallel magnetization alignment between the top and bottom magnetic TI layers. For simplicity, we set $\hbar v_{\mathrm{F}}$=1, $m_1$= 1, $m_0$= -0.5 and $|m_{\mathrm{z}}(\mathrm{r})|$=3. Figures 4e and 4f plot a local Chern marker [65,66] to visualize the local topological feature in real space. Our calculations show that the percentage of the $C$=0 magnetic domains increases as the percentage of antiparallel magnetization states $n_0$ increases (Figs. 4e, 4f, and S7). Here $n_0 = N_0/N_T$ with $N_0$ and $N_T$ being the numbers of antiparallel and total magnetic domains, respectively. Both $N_0$ and $N_T$ are dependent on the thickness $m$ (Fig. S8). Figure 4d plots the renormalized localization length $\Lambda \equiv \lambda/L$ as a function of $n_0$ for different sizes ($L$=48 ~128). Here, $\lambda(L)$ is the localization length of the cylinder sample of circumference $L$ [68]. A transition from a CM phase with $d\Lambda/dL$=0 to an insulator phase with $d\Lambda/dL$<0 occurs as the percentage of the antiparallel magnetization (i.e., $C$=0) domains $n_0$ increases. This transition is driven by the cessation of percolation in the chiral edge modes around magnetic domains, resulting from an insufficient presence of $C$=±1 domains. Our prior study [61] has predicted a similar CMIT, where the disconnection of the chiral edge modes is driven by an increase in disorder strength rather than the formation of antiparallel magnetization (i.e., $C$=0) domains. Therefore, the model Hamiltonian simulation supports our interpretation that IEC alters the number of $C$=0 magnetic domains and, in turn, leads to the occurrence of the CMIT in our QAH insulators. Moreover, since the averaged Chern number of the entire system consistently remains zero [58], the CMIT observed here differs significantly from the conventional QH PPT, which involves a change in the Chern number. This phenomenon is reminiscent of the CMIT predicted in 2D electron gas subjected to random magnetic fields [50,58,59] and has recently renewed interest in topological systems [50,52,58-63].

Our theoretical model reveals that the proportion of antiparallel magnetic domains influenced by the IEC can lead to a CMIT from $m$=4 to $m$=20 near QAH PPT, where the IEC is mainly affected by $m$. The nearly temperature-independent behavior of the $\rho_{xx,\max}$ values in the $m$=4 and $m$=8 sandwiches signifies a CM phase in the strong IEC limit. For $m$=20, the IEC strength decreases, leading to an $m$ change-induced CMIT. Furthermore, a temperature-driven MIT occurs for an intermediate thickness $m$=16, likely due to enhanced scattering caused by increased randomness from thermal spin fluctuations in the magnetic degree of freedom as *T* increases [69]. As *m* increases from 20 to 100, the dominant conductance shifts from the top and bottom surfaces to the side surfaces, decreasing $\rho_{xx,\max}$ [33,43]. Here, the conductance of the top and bottom surfaces and the side surfaces is assumed to be independent. In addition, the CM phase is absent in the *m*=0 sample and weak in the *m*=1 sample (Fig. S3) [37]. These observations suggest that the CM phase depends on the quantization and uniformity of the QAH insulators.

To summarize, we realize the well-quantized QAH effect in magnetic TI sandwiches with the TI spacer layer thickness *m* from 1 to 100. We find that varying *m* can lead to the occurrence of an IEC-induced CMIT near QAH PPT. Thinner QAH samples exhibit 2D CM behaviors near PPT, while thicker QAH samples manifest as 3D insulators. The IEC-induced CMIT near QAH PPT is well explained through a magnetic TI Hamiltonian with random magnetic domains. Our findings provide insights into the interplay between the top and bottom surface states in QAH insulators and will motivate further investigation into manipulating quantum phase transitions via layer thickness.

**Acknowledgments:** We thank Z. Bi and D. Xiao for their helpful discussions. This work is primarily supported by the ONR Award (N000142412133), including MBE growth, dilution transport measurements, and theoretical calculations. The sample characterization is supported by the NSF grant (DMR-2241327). C. -Z. Chang acknowledges the support from the Gordon and Betty Moore Foundation's EPiQS Initiative (GBMF9063 to C. -Z. C.). C.-Z. Chen acknowledges the support from the National Key R&D Program of China Grant (2022YFA1403700), the Natural Science Foundation of Jiangsu Province Grant (BK20230066), and the Jiangsu Shuang Chuang Project (JSSCTD202209).

**Data availability:** The data that support the findings of this article are openly available [70], embargo periods may apply.

**Figures and figure captions:**

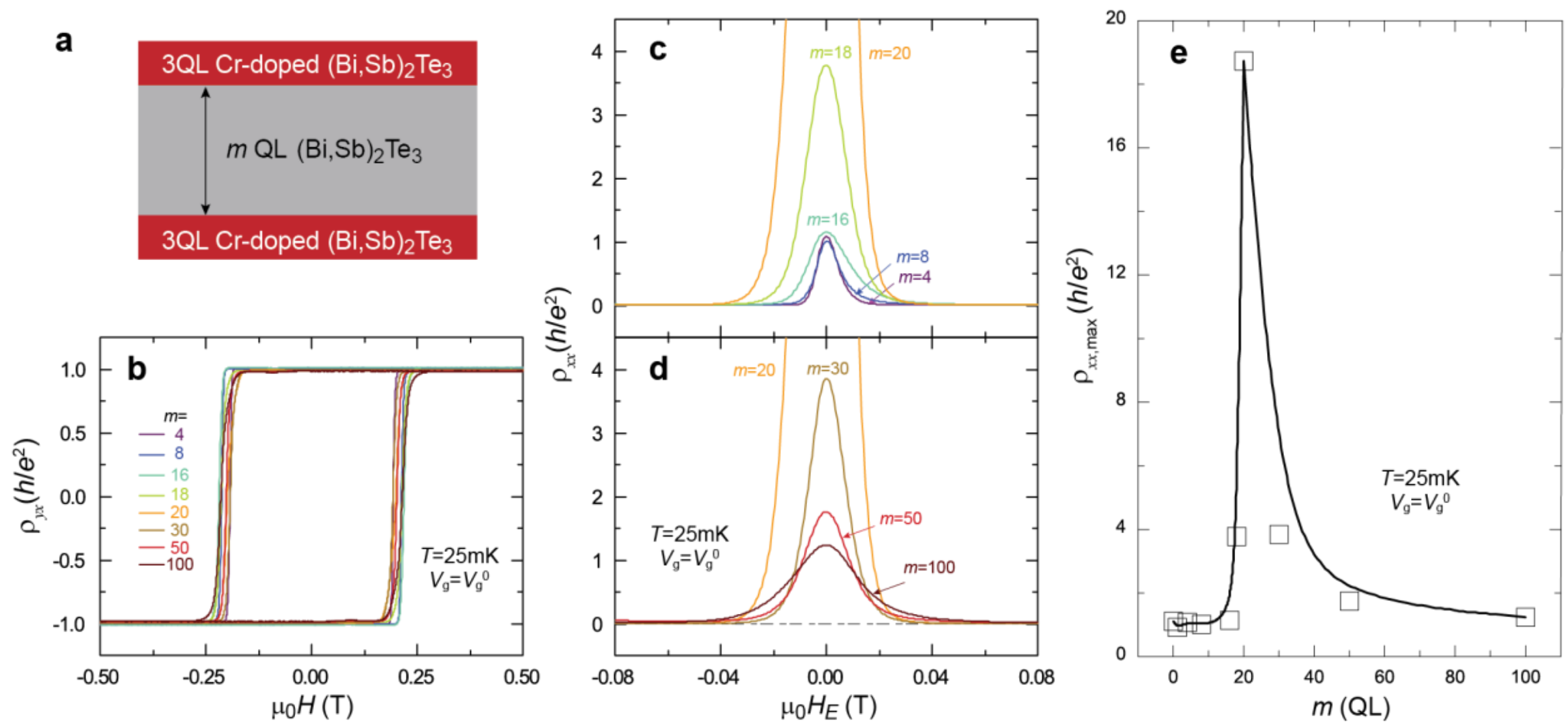

**Fig. 1| The QAH state in magnetic TI sandwiches with different *m*. a**, Schematic of the magnetic TI sandwich. **b**, $\mu_0 H$-dependent $\rho_{yx}$ of the samples with $4 \leqslant m \leqslant 100$. **c, d,** The effective magnetic field $\mu_0 H_E$ dependent $\rho_{xx}$ of the samples with $4 \leq m \leq 20$ (**c**) and $20 \leq m \leq 100$ (**d**). **e,** $\rho_{xx,\ max}$ near $\mu_0 H_c$ as a function of $m$. All measurements are taken at $V_g = V_g^0$.

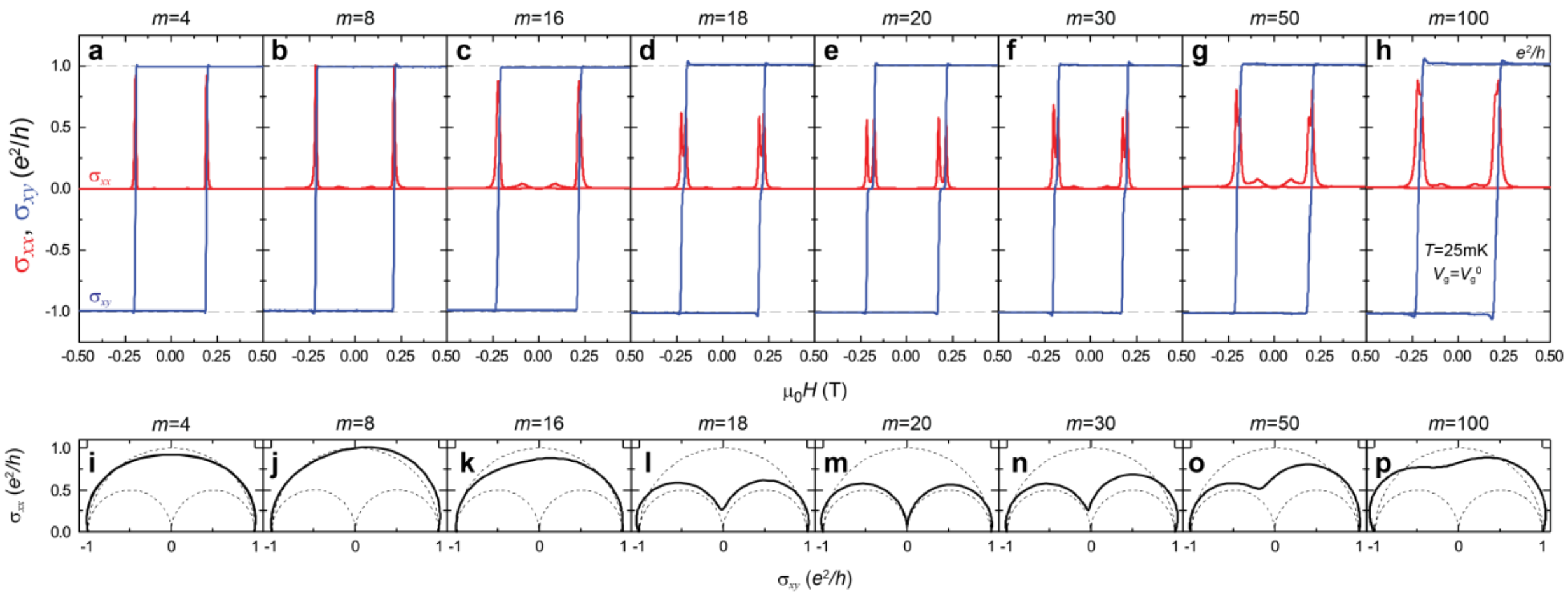

**Fig. 2| Flow diagrams of magnetic TI sandwiches with different *m*. a-h**, $\mu_0 H$-dependent $\sigma_{xx}$ (red) and $\sigma_{xy}$ (blue) of the samples with *m*=4 (**a**), *m*=8 (**b**), *m*=16 (**d**), *m*=18 (**e**), *m*=20 (**e**), *m*=30 (**f**), *m*=50 (**g**), and *m*=100 (**h**). **i-p**, Flow diagrams of ($\sigma_{xy}$, $\sigma_{xx}$) of the samples with *m*=4 (**i**), *m*=8 (**j**), *m*=16 (**k**), *m*=18 (**l**), *m*=20 (**m**), *m*=30 (**n**), *m*=50 (**o**), and *m*=100 (**p**). Two semicircles of radius $e^2/2h$ centered at ($e^2/2h$, 0) and ($-e^2/2h$, 0) and one semicircle of radius $e^2/h$ centered at (0, 0) are shown in dashed lines. All measurements are taken at $V_g = V_g^0$ and $T$=25mK.

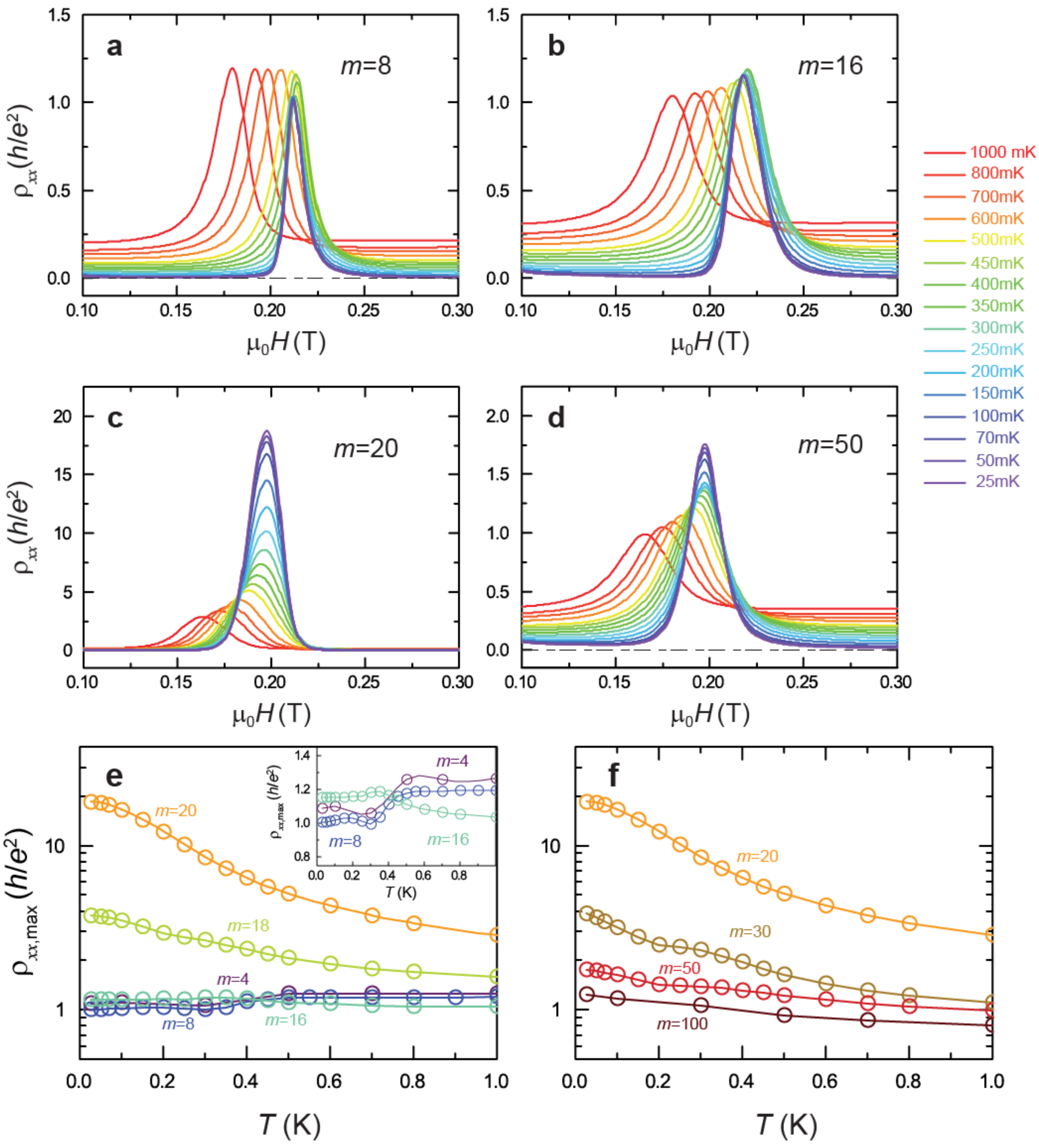

**Fig. 3| IEC-induced CMIT in QAH sandwiches. a-d,** $\mu_0 H$-dependent $\rho_{xx}$ of the samples with *m*=8 (**a**), *m*=16 (**b**), *m*=20 (**c**), and *m*=50 (**d**) near PPT at different temperatures. **e, f**, Temperature-dependent $\rho_{xx,\,max}$ with different *m*. Inset of (**e**): Enlarged $\rho_{xx,\,\mathrm{max}}$ -*T curves of the m*=4*, m*=8, and *m*=16 samples. All measurements are taken at $V_g$=$V_g^0$.

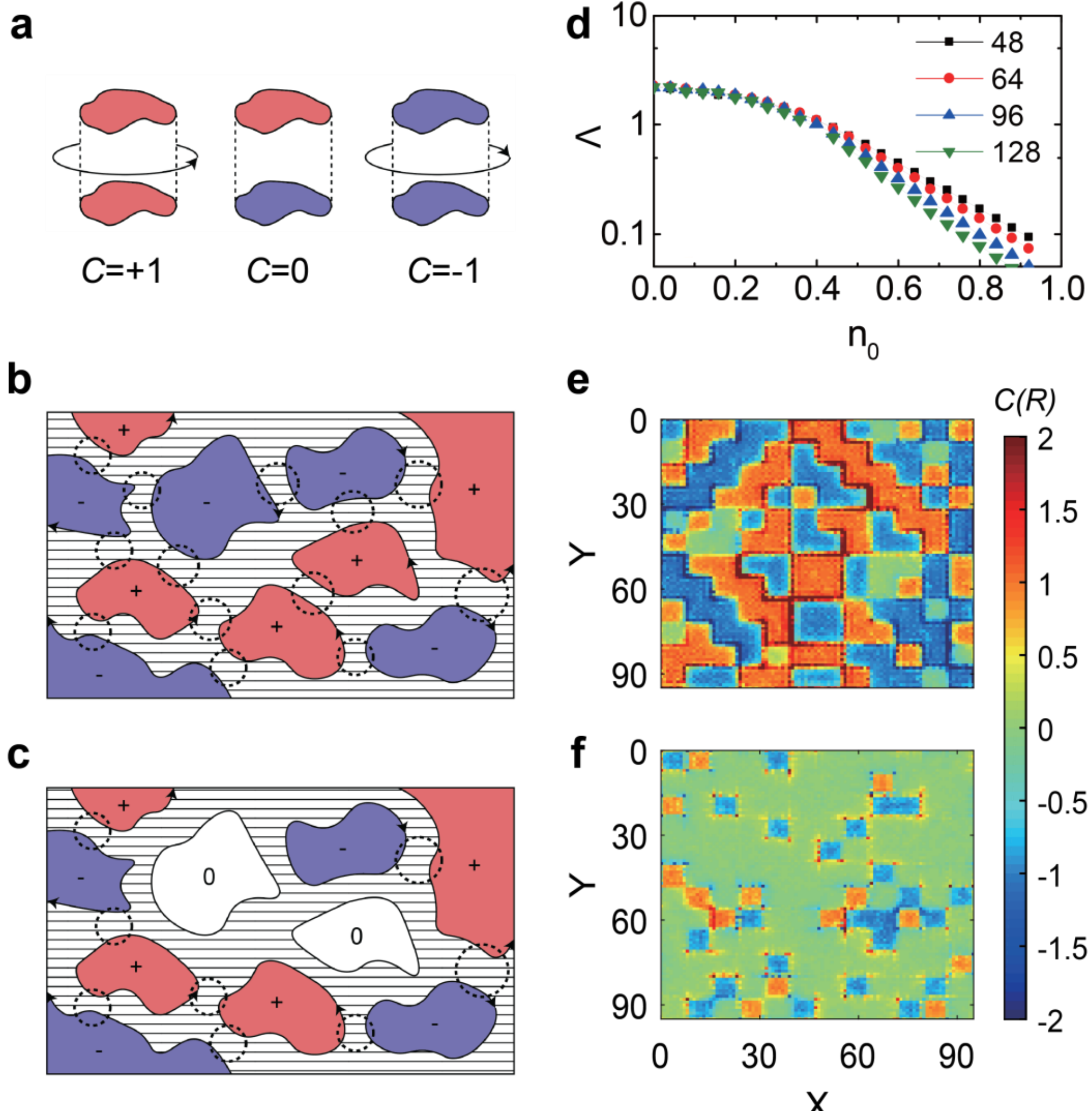

**Fig. 4| Theoretical calculations of IEC-induced CMIT in QAH sandwiches. a,** Chern number $C$ and chiral edge modes of magnetic domains with parallel and antiparallel magnetization alignment. **b, c**, Magnetic domains distribution near PPT. "+", "–", and "0" denote $C$=1, $C$= -1, and $C$=0 magnetic domains, respectively. The dashed circles indicate the tunneling between magnetic domains. **d,** Renormalized localization length Λ as a function of the antiparallel magnetization domain density $n_0$ for different sizes ($L$=48~128). **e, f,** Topography (96 × 96 cells) of local Chern maker *C(R)* of the magnetic TI Hamiltonian under $n_0 = 0.1$ (**e**) and $n_0 = 0.7$ (**f**). $n_0$ is defined as $n_0$=$N_0/N_T$. Here $N_0$ is the number of the antiparallel magnetization (i.e., $C$ =0) domains and $N_T$ is the total number of magnetic domains.